\documentclass[proof]{WileyASNA-v1}

\usepackage{graphicx}
\usepackage{tensor}
\usepackage{verbatim}

\newcommand{\quarter}{{\textstyle\frac{1}{4}}}
\newcommand{\onehalf}{{\textstyle\frac{1}{2}}}
\newcommand{\onethird}{{\textstyle\frac{1}{3}}}
\newcommand{\threequarter}{{\textstyle\frac{3}{4}}}
\newcommand{\fref}[1]{Fig. \ref{#1}}

\articletype{Article Type}%

\received{18 November 2020}
\revised {18 November 2020}
\accepted{18 November 2020}

\begin{document}

\title{Rigorous derivation of dark energy and inflation as geometry effects in Covariant Canonical Gauge Gravity}

\author[1]{David Vasak*}
\author[1]{Johannes Kirsch}
\author[1,2]{J\"urgen Struckmeier}
\authormark{D. Vasak et al.}

\address[1]{\orgname{Frankfurt Institute for Advanced Studies (FIAS)}, \orgaddress{Ruth-Moufang-Str. 1, 60438~Frankfurt am Main}, \country{Germany}}
\address[2]{\orgname{Goethe Universit\"at}, \orgaddress{Max-von-Laue-Str. 1, 60438~Frankfurt~am~Main}, \country{Germany}}
\corres{*\email{vasak@fias.uni-frankfurt.de}}

\abstract{
The cosmological implications of the Covariant Canonical Gauge Theory of Gravity (CCGG) are investigated. CCGG is a Palatini theory derived from first principles using the canonical transformation formalism in the covariant Hamiltonian formulation. The Einstein-Hilbert theory is thereby extended by a quadratic Riemann-Cartan term in the Lagrangian. Moreover, the requirement of covariant conservation of the stress-energy tensor leads to necessary presence of torsion. In the Friedman universe that promotes the cosmological constant to a time-dependent function, and gives rise to a geometrical correction with the EOS of dark radiation. The resulting cosmology, compatible with the $\Lambda$CDM parameter set, encompass)es bounce and bang scenarios with graceful exits into the late dark energy era. Testing those scenarios against low-$z$ observations shows that CCGG is a viable theory.}


\maketitle

\section{Introduction} \label{sec:Intro}

The motivation for this work is to explore the potential of the novel Covariant Canonical Gauge Gravity (CCGG) and the hope to shed new light on some of the mysteries of standard cosmology. That cosmology is based on Einstein's General Relativity, a phenomenology-driven theory created by Einstein. Later, concepts like dark matter, dark energy, or inflation, have been added to close substantial gaps to observations, which are yet lacking agreed physical understanding. CCGG is, in contrast, based on a rigorous mathematical framework that is rooted in just a few fundamental assumptions~(\citep{struckmeier08, struckmeier13, struckvasak15, struckmeier17a, struckmeier18a, struckvasak18c}).

\medskip
In this paper we present the results of a first, preliminary analysis of the CCGG-Friedmann universe focussing on selected low-$z$ observations. It is organized as follows.
We start by briefly sketching the philosophy and relevant features of CCGG. Considering gravity as a gauge field is not new~\citep{utiyama56, kibble67, sciama62, hehl76, hayashi80} but here we rely on the mathematical rigorousness of the canonical transformation theory in the de Donder-Hamiltonian formulation~\citep{dedonder30}. This framework naturally yields a Palatini (first-order) theory in the Riemann-Cartan geometry. Torsion and a quadratic Riemann-Cartan term are new ingredients modifying the Einstein-Hilbert ansatz for vacuum gravity. As discussed in Refs. \citep{vasak19a, vasak19b} CCGG does not need to invoke any ad hoc higher-order curvature terms and/or auxiliary scalar fields~\citep{wetterich88, wetterich15, starobinsky80, starobinsky82} to generate interesting scenarios of cosmological evolution.

\medskip
In an isotropic (Friedman) universe\citep{friedman22}, filled with homogeneous (standard) matter components approximated by ideal fluids, the cosmological constant is promoted to a cosmological field, and the quadratic extension gives rise to a \emph{geometrical stress} tensor with the character of dark radiation. The dynamical cosmological term arises due to the presence of torsion, and dark energy appears as an energy reservoir based on a local~contortion density. (For further discussions on torsion and cosmology see also~\citep{capozziello02, capozziello03, capozziello14, chen09, hammond02, minkevich07, minkowski86, shie18, unger19}. Numerical results are presented sketching cosmological scenarios arising from the interplay of these ``dark'' components.

\section{Covariant Canonical Gauge Gravity} \label{sec:CCGG}

The canonical approach to gauge gravity emanates from several key principles:

\begin{enumerate}
 \item Principle of Least Action: The dynamics of of the classical field theory of matter and curvilinear spacetime geometry derives via variation from an action integral which is a world scalar.
 \item Equivalence Principle: A local inertial (Minkowski) frame must exist at any point of the space-time manifold that is defined up to local Lorentz transformations.
 \item Principle of General Relativity: The dynamics of the system must be invariant with respect to arbitrary coordinate transformations (diffeomorphisms).
 \item Principle of Information Conservation: The integrand of the action integral, the Lagrangian density, must be reversibly (Legendre) transformeable into Hamiltonian densities\footnote{This is not necessary but sufficient, see \citep{smetanova18}.}, i.e. \emph{non-degenerate} or \emph{regular}.
\end{enumerate}

Einstein's Principle of General Relativity and the Equivalence Principle relevant for gravity are incorporated by a ("Lorentzian") frame bundle with fibers spanned by ortho-normal bases fixed up to arbitrary (local orthochronous) Lorentz transformations. The gauge group underlying the CCGG approach is thus the SO($1,3$)$^{(+)}\, \times \,$Diff$(M)$ group. The covariant canonical transformation theory then implements form-invariance of the action integral with respect to that gauge group\footnote{Struckmeier et al.  \citep{struckmeier08, struckmeier13} have, as a proof of concept, derived the Yang-Mills gauge theory from first principles.}
without any detour to $1+3$ splitting or the Dirac formalism. The (spin) connection coefficients emerge thereby as the  gauge fields.
The gauge field is independent of the metric tensor (or vierbein fields) which are fundamental structural elements of the Lorentzian manifold. Moreover, the postulated regularity of the Lagrangian implies that it must contain an at least quadratic Riemann-Cartan tensor concomitant \citep{benisty18a}. The quadratic term, controlled by a new dimensionless \emph{deformation} parameter, is therefore chosen as the minimal extension of Einstein's linear Lagrangian.     In this way the framework delivers a classical, linear-quadratic, first-order (Palatini) field theory where the connection and metric mediate gravitation. The couplings of matter fields and gravity are unambiguously fixed, and space-time is endowed with kinetic energy and inertia\footnote{Since this is a Palatini theory, the Ostrogradsky instability theorem does not apply. \citep{ostrogradsky1850, woodard15}.}. The resulting space-time geometry is not a priori constrained to zero torsion and/or metric compatibility, may nevertheless implement these restrictions dynamically via canonical equations of motions.

\medskip
\!The so called consistence equation in CCGG is a combination of the Euler-Lagrange equations extending the field equation of GR:
\begin{align}
\label{eq:modEinstein}
  -{\Theta}_{\mu\nu} :=& \,g_1\left( R_{\alpha\beta\gamma\mu}\, {R}\indices{^{\alpha\beta\gamma}_\nu}
           - \quarter g_{\mu\nu} \, {R}_{\alpha\beta\gamma\delta}\, {R}^{\alpha\beta\gamma\delta}\right) \\
           &+ \frac{1}{8\pi G}\, \left[ {R}\indices{_{(\mu\nu)}}
           - g_{\mu\nu} \left( \onehalf {R} + \lambda_0 \right)\right] = {T}\indices{_{(\mu\nu)}}. \nonumber
\end{align}
Interpreting ${\Theta}_{\mu\nu}$ on the l.h.s. as the energy-momentum (``strain'') tensor of space-time, this equation can be interpreted as a balance equation between the strain-energy and the stress-energy tensor ${T}\indices{_{(\mu\nu)}}$ of matter.
The dimensionless coupling constant $g_1$ controls the admixture of quadratic gravity to GR, $G$ is Newton's gravitational constant, and we call $\lambda_0$ the "bare" cosmological constant. The Riemann-Cartan tensor,
\begin{equation*}
R\indices{^{\alpha}_{\beta\mu\nu}} =
 \gamma\indices{^{\alpha}_{\beta\nu,\mu}} -
 \gamma\indices{^{\alpha}_{\beta\mu,\nu}} +
 \gamma\indices{^{\alpha}_{\xi\mu}}\gamma\indices{^{\xi}_{\beta\nu}} -
 \gamma\indices{^{\alpha}_{\xi\nu}}\gamma\indices{^{\xi}_{\beta\mu}}
 \end{equation*}
is in general built from an asymmetric connection, and the symmetric portion of the stress-energy tensor is the source term on the r.h.s. of the field equation. (The conventions $(+,\,-,\,-,\,-)$ for the metric signature and natural units $\hbar = c = 1$ are applied. A comma denotes partial derivative, and indices in (brackets) parentheses indicate (anti-)symmetrization.

\section{Geometrical stress energy and Cartan contortion density} \label{sec:CovCons}
In this paper we wish to explicitly work out the differences invoked by the CCGG model to the standard, GR based so called $\Lambda$CDM cosmology, and hence assume here both, a torsion-free geometry \emph{and} the stress-energy tensor to be covariantly conserved,
$ \bar{\nabla}_\nu \bar{T}\indices{^{(\mu\nu)}} = 0.$
(Here and in the following quantities based on the torsion-free Levi-Civita connection $\bar{\gamma}\indices{^{\lambda}_{\mu\nu}} =
\genfrac{\lbrace}{\rbrace}{0pt}{0}{\lambda}{\mu\nu}$ are marked by a bar.) This is, however, inconsistent with the behaviour of the strain-energy tensor as in general $ \bar{\nabla}_\nu \bar{\Theta}\indices{^{(\mu\nu)}} \ne 0$. This can readily be seen:
Defining the (symmetric and traceless) quadratic (Kretschmann) concomitant
\begin{equation}
 Q\indices{^\mu^\nu} :=  R^{\alpha\beta\gamma\mu}\, R\indices{_{\alpha\beta\gamma}^{\nu}}
           - \quarter g\indices{^\mu^\nu} \, R^{\alpha\beta\gamma\xi}\, R_{\alpha\beta\gamma\xi}
\end{equation}
and the (symmetric) Einstein tensor
\begin{equation}
G\indices{^\mu^\nu} := R\indices{^{(\mu\nu)}} - \onehalf g\indices{^\mu^\nu} \, R
\end{equation}
we find
\begin{align}
\bar{\Theta}\indices{_{\nu}^{\mu}_{;\mu}} &= \bar{Q}\indices{_{\nu}^{\mu}_{;\mu}}
= \bar{R}\indices{_{\alpha\beta\gamma}^\nu} \, \bar{\nabla}_\mu \bar{R}\indices{^{\alpha\beta\gamma\mu}} \label{covderQ} \\
\bar{G}\indices{_{\nu}^{\mu}_{;\mu}} &\equiv 0.
\end{align}
Rather than being a vanishing identity as it is for the Einstein tensor, the expression on the r.h.s. of Eq.~(\ref{covderQ}) gives a relation between metric and connection. If for a specific ansatz for the metric the condition
\begin{equation} \label{covderR}
\bar{R}\indices{_{\alpha\beta\gamma}^\nu} \, \bar{\nabla}_\mu \bar{R}\indices{^{\alpha\beta\gamma\mu}}  = 0
\end{equation}
is violated, we obviously have to abandon the Levi-Civita connection and accept an asymmetric connection. It is well known that in metric compatible space-times
this means
\begin{equation}\label{eq:gammaing}
\gamma\indices{^{\lambda}_{\mu\nu}} =
\genfrac{\lbrace}{\rbrace}{0pt}{0}{\lambda}{\mu\nu}+
K\indices{^{\lambda}_{\mu\nu}}
\end{equation}
where $K\indices{_{\lambda}_{\mu\nu}} \,=\,
S\indices{_{\lambda\mu\nu}} - S\indices{_{\mu\lambda\nu}} + S\indices{_{\nu\mu\lambda}}\,=\,-K\indices{_{\mu}_{\lambda\nu}}$
is the contortion tensor, a combination of metric and the Cartan torsion tensor $S\indices{^\lambda_{\mu\nu}} = \onehalf(\gamma\indices{^{\lambda}_{\mu\nu}}- \gamma\indices{^{\lambda}_{\nu\mu}})$. Invoking torsion is thus \emph{necessary} in this case for the condition \eqref{covderR} to hold.

\medskip
The terms in Eq.~(\ref{eq:modEinstein}) that modify Einstein's field equation due to the quadratic terms and torsion can now be explicitly worked out. The Riemann-Cartan tensor
\begin{equation}
 R\indices{_{\alpha\beta\gamma\sigma}}(\gamma\indices{^{\lambda}_{\mu\nu}})
 \equiv \bar{R}\indices{_{\alpha\beta\gamma\sigma}} + P\indices{_{\alpha\beta\gamma\sigma}},
\end{equation}
separates into the Riemann tensor commanding the Levi-Civita connection, and the torsion-related correction, the \emph{Cartan curvature tensor}
\begin{equation} \label{def:Ptensor}
P\indices{_{\lambda}_{\sigma\mu\nu}}
 := \bar{\nabla}_\mu K\indices{_{\lambda}_{\sigma\nu}} -
    \bar{\nabla}_\nu K\indices{_{\lambda}_{\sigma\mu}} -
    K\indices{_{\lambda}_{\beta\nu}} K\indices{^{\beta}_{\sigma\mu}} + K\indices{_{\lambda}_{\beta\mu}} K\indices{^{\beta}_{\sigma\nu}}.
\end{equation}
Similarly, the Einstein tensor becomes
\begin{equation}
G\indices{^\mu^\nu} = \bar{G}\indices{^\mu^\nu} + P^{(\mu\nu)}
           - \onehalf \,g^{\mu\nu}\,  P,
\end{equation}
and Eq.~(\ref{eq:modEinstein}) can be brought into the ``Einstein form''
\begin{equation} \label{eq:modEinstein5}
             \frac{1}{8\pi G}\,\left[
           \bar{G}^{\mu\nu} - g^{\mu\nu} \Lambda(x) \,\right]
           = \bar{T}\indices{^{(\mu\nu)}} - g_1\, {Q}^{\mu\nu}
           - \frac{1}{8\pi G} P'^{(\mu\nu)}
,
\end{equation}
which for $g_1 = 0$ and $S\indices{^\lambda_{\mu\nu}} = 0$ coincides with Einstein's field equation. Notice that for an application in cosmology with just classical matter we will neglect the spin-torsion interaction by assuming the stress-energy tensor of matter to be independent of the affine connection and hence of torsion, giving $\bar{T}\indices{^{(\mu\nu)}} \equiv {T}\indices{^{(\mu\nu)}}$. Furthermore, $P'^{(\mu\nu)}:=P^{(\mu\nu)} - \quarter g^{\mu\nu} P $
is the trace-free Cartan-Ricci tensor which as all tensors in this equation is symmetric by definition, including the quadratic Riemann concomitant~$Q^{\mu\nu}$.
The Cartan-Ricci curvature scalar $P(x)$ built from contortion and metric promotes the cosmological constant to the \emph{cosmological function}
\begin{equation} \label{def:cosmfield}
  \Lambda(x) := \lambda_0 - \quarter P(x).
\end{equation}
representing \emph{dark energy}\footnote{As shown in \citep{vasak19b} the bare cosmological constant acquires in this theory a contribution from that quadratic term to the vacuum energy denoted $g_3$, giving $\lambda_0 = 3/16\pi G\,g_1 +8\pi G \,g_3$. This reliefs the identification of the cosmological constant with the vacuum energy shedding on light on the cosmological constant problem.}.

\medskip
The geometric tensor corrections, now moved to the r.h.s. of the CCGG consistency  equation, appear as a new, trace-free geometrical stress-energy tensor representing \emph{dark radiation}. This is justified as they are trace-free in analogy to the energy-momentum tensor of radiation or relativistic matter. This re-arrangement enables now to study the newly emerging phenomena of dark energy and dark radiation in relation to General Relativity in a standard cosmological model.


\section{CCGG cosmology} \label{sec:Friedman}

To align with the Cosmological Principle of a homogeneous and isotropic universe the Friedman-Lema\^{i}tre-Robertson-Walker (FLRW) metric
\begin{equation}\label{def:RWmetric}
ds^2 = dt^2 + a^2(t) \left[\frac{dr^2}{1-K_0 r^2} - r^2\left(d\theta^2 + \sin^2(\theta)\,d\varphi^2\right)\right].
\end{equation}
is assumed to describe the space-time geometry.
The dimensionless scale parameter $a(t)$ is a function of the cosmological time $t$ and the only dynamical freedom of the theory. It is normalized such that $a(t_0) = 1$ applies to today, i.e. to time $t_0$. The parameter $K_0$ distinguishes between three fundamental geometry types: $K_0 = 0$ flat, $K_0 > 0$ spherical, $K_0 < 0$ hyperbolic.

\medskip
Calculating now the Christoffel symbols and the curvature tensors we find Eq.~(\ref{covderR}) violated. Hence,~Eq.~(\ref{eq:modEinstein5}) must be considered with the tensor corrections as outlined above.  The torsion tensor must be selected such that it ensures the covariant conservation of the strain-energy tensor. This will be left to a future investigation and we perform a first analysis by neglecting the torsion dependent stress tensors. We thus substitute ${Q}^{\mu\nu} = \bar{Q}^{\mu\nu}$ and retain torsion only in the cosmological field as a novel dynamical quantity~$\Lambda(x)$. In this geometry that dark energy term
can only depend on the universal time~$t$. The analysis is further simplified by adopting the scaling ansatz
\begin{equation} \label{def:f(a)}
\Lambda(t) = \Lambda(t(a)) =: \Lambda_0 \, f(a)
\end{equation}
with the dimensionless function $f(a)$ and  a constant~$\Lambda_0$ which is a parameter equivalent to the $\Lambda$ of the $\Lambda$CDM ansatz. Plugging this into the CCGG consistency equation \eqref{eq:modEinstein5} gives the Friedman equation
 \begin{align} \label{eq:modFriedman4}
 \frac{H^2(a)}{H_0^2}
  &=
 \sum_{i=r,m, K}\,\Omega_i\,a^{-n_i} + \Omega_\Lambda\,f(a) + \Omega_g(a),
\end{align}
where $H(a)$ is the Hubble function, $H_0 \equiv H(1)$.
The constants $\Omega_i$ are identical to the $\Lambda$CDM density parameters
\begin{equation} \label{scalinglaw1}
 \Omega_i := \frac{8\pi G}{3H_0^2}\,\rho_i\,a^{n_i} = \mathrm{const.}, \qquad i=r, m.
\end{equation}
and
\begin{subequations} \label{scalinglaw2}
\begin{alignat}{4}
   \Omega_K &:= -\frac{K_0}{H_0^2} \qquad \,\omega_K = -\onethird, \,n_K = 2   \label{def:Ck}\\
   \Omega_\Lambda &:= \frac{\Lambda_0}{3H_0^2}  \qquad \omega_\Lambda = -1, \,n_\Lambda = 0, \label{def:CLambda} \\
\end{alignat}
\end{subequations}
with $\omega_i, i = m,r,...$ denoting the equation of state, and $n_i$ the scaling property of the density of matter, radiation etc. $\Omega_g(a)$ represents the geometrical effects emerging from the quadratic term \citep{vasak19a}:
\begin{equation} \label{def:chi}
 \Omega_g(a) :=
 \frac{
 \left( \quarter \Omega_m  a^{-3} + \Omega_\Lambda \,f(a) \right) \left( \threequarter \Omega_m \, a^{-3} + \Omega_r\,a^{-4} \right)
 }
 {
 \onehalf g_2 - \quarter \Omega_m\,a^{-3} - \Omega_\Lambda \,f(a)
 }.
 \end{equation}
where for convenience we use
\begin{equation} \label{def:g2}
g_2 := \frac{1}{16 \pi G g_1 H_0^2}.
\end{equation}
$\Omega_g(a)$ is well defined since the function $f(a)$ obeys the unique differential equation \citep{vasak19b}
\begin{subequations} \label{ODE:f(a)}
\begin{alignat}{4}
 \frac{df}{da} &=
  \frac{3\Omega_m}{4\Omega_\Lambda} \,\frac{A(a) - B(a) \left(\quarter \Omega_m a^{-3} + \Omega_\Lambda \,f(a) \right)}{a^4 \left( A(a) + B^2(a) \right)} \\
 A(a) &=: \onehalf g_2 \left(\threequarter \Omega_m a^{-3} + \Omega_r a^{-4}\right) \\
 B(a) &=: \onehalf g_2 - \quarter \Omega_m a^{-3} - \Omega_\Lambda \,f(a) 
\end{alignat}
\end{subequations}

\begin{figure}[htb]
\includegraphics[width=70mm]{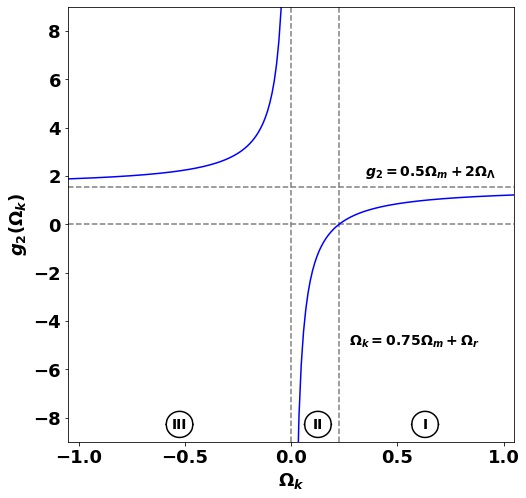}
\caption{\footnotesize
As $g_2$ must be non-zero and finite the root $\Omega_K = - 3\Omega_m / 4 +  \Omega_r$, where $g_2 = 0$, and $\Omega_K = 0$ where $g_2 = \pm\infty$, are both ``forbidden'' values. This divides the parameter space in three disjoint Regions denoted by I, II and III with different combinations of sign$(\Omega_K)$ and sign$(g_2)$. For $\Omega_K  \rightarrow \pm\infty$ the coupling constant converges to $g_2  = \Omega_m/2 + 2\Omega_\Lambda$.}
\label{fig:g_2(K_0)}
\end{figure}

\medskip
We now require that the dark energy term coincides with the observed present-day value of the cosmological constant. Setting $\Lambda_0 = \Lambda_{\mathrm{obs}}$ gives then the initial condition~$f(1)=1$, and Eq.~(\ref{eq:modFriedman4}) reduces to
 \begin{equation} \label{eq:modFriedman5}
1 =  \sum_{i=r,m, \Lambda,K,g}\,\Omega_i.
\end{equation}
\begin{table}[ht]
 \begin{tabular}[t]{|c|c|c|c|c|} \hline \hline
Data & $\Omega_\Lambda$ & $\Omega_m$ & $\Omega_r$ & $h_0$\\
\hline \hline
Default & $0.69990$ & $0.30000 $ & $0.00005$ & $0.70903$
\\
\midrule
\textcolor{black}{Late} & $0.70000$ & $0.30000 $ & $0.00005$ & $0.74500  $
\\
\midrule
Early & $0.68500$ & $0.31500 $ & $0.00005$ & $0.67400  $
\\
\hline \hline
\end{tabular}
\caption{\footnotesize
The $\Lambda$CDM parameter sets used for the sensitivity check of the Hubble diagram fit. The data are taken from the Refs. \citep{planck15} (= Default, applied throughout this paper), \citep{dhawan20} (Late) and \citep{planck18} (Early).
}
\label{tab:Parameter2}
\end{table}
In order to align the parameters with the flat $\Lambda$CDM or Concordance Model
 for which $\sum_{i=r,m, \Lambda}\,\Omega_i = 1$, the curvature and the geometry terms must just cancel each other:
 \begin{equation} \label{eq:k0}
-\Omega_K = \Omega_g =
 \frac{
 \left( \quarter \Omega_m  + \Omega_\Lambda \right) \left( \threequarter \Omega_m + \Omega_r \right)
 }
 {
 \onehalf g_2 - \quarter \Omega_m - \Omega_\Lambda
 }.
 \end{equation}
This relation can be resolved for $g_2$,
 \begin{equation} \label{prooffne1}
  \onehalf \, g_2(\Omega_K) = \frac{1}{\Omega_K} \, \left(
  \quarter\,\Omega_m+\Omega_\Lambda\right) \left(\Omega_K - \threequarter \,\Omega_m -\Omega_r  \right),
 \end{equation}
and is visualized in~Fig. \ref{fig:g_2(K_0)}. (By Eq.~(\ref{def:g2}) this can easily be transformed into a relation of the curvature parameter $\Omega_K$ and the deformation parameter $g_1$ of the theory.)
By this arrangement of the constants, $g_2$ diverges with $\Omega_K \rightarrow 0$, and then of course $g_1$ approaches zero. However, while the limiting case $g_1 = 0$ seems to recover the Hubble equation of standard cosmology, the limiting process,~$g_1 \rightarrow 0$, is continuous but not convergent\footnote{Neither $g_1$ nor $g_2$ can be continuously connected to the value $0$ as then the quadratic term in either the Hamiltonian or in the Lagrangian would diverge.}.
With $g_1$ and $g_2$ finite $\Omega_K = 0$ is excluded making the set of solutions non-compact.

\medskip
For later use we note that the Friedman equation~(\ref{eq:modFriedman4}) can be re-written as an equation of motion of a classical fictitious point particle with the dimensionless mass $2$ in an external potential $V(a)$:
\begin{align}
 &\dot{a}^2 + V(a) = H_0^2\,\Omega_K \label{DG1}\\
&V(a) = -H_0^2\left[ \Omega_r\,a^{-2} + \Omega_m\,a^{-1} + \Omega_\Lambda\,a^{2}f(a) + \Omega_g(a)\,a^{2} \right]. \label{def:V(a)}
\end{align}
The particle's kinetic energy is $\dot{a}^2 \equiv H^2(a)\, a^2$, and its total energy $H_0^2\,\Omega_K = -K_0$.

\medskip
An important astronomical observable is also the dimensionless deceleration function
\begin{equation} \label{eq:deceleration}
q := -\frac{\ddot{a}}{\dot{a}^2}\,a \equiv -\frac{\ddot{a}}{a}\,\frac{1}{H^2(a)}
,
\end{equation}
which explicitly depends on the curvature parameter $K_0$, and implicitly on the dark energy and curvature functions in the Hubble function.
For the $\Lambda$CDM ``Default'' parameter set (cf. Table~1) the present-day deceleration parameter~$q_0 \equiv q(1)$ is
\begin{equation}\label{eq:decelerationf0}
q_0
 \approx -0.55 + K_0/H_0^2.
\end{equation}
The values of both, $\Omega_K = -K_0/H_0^2$ and $g_2$, are thus restricted by the measurement accuracy of $q_0$ \citep{planck15, bernal17, camarena20}.

\medskip
In order to test the viability of the CCGG-Friedman model within the present accuracy of observations we conduct a preliminary analysis with the four priors of the flat Concordance Model and focus on investigating the influence of the CCGG deformation parameter $g_1$. Some key results of our numerical analysis are presented in the next section.
\begin{figure}[htb]
\includegraphics[width=70mm]{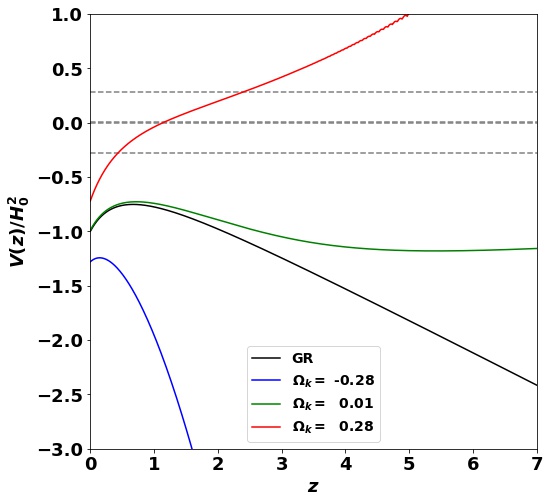}
\caption{\footnotesize The $z$- dependence of the scale potentials $V(z;g_2(\Omega_K))/H_0^2$ with the values $\Omega_K = 0.28, 0.01, -0.28$ typical for the Regions I, II and III, respectively. In Region I a turning point arises where the potential crosses the line $\Omega_K = 0.28$. Potentials of Regions II and III do not cross the corresponding $\Omega_K$-lines at $-0.28$ and $0.01$. The curve labeled GR shows the potential of the standard $\Lambda$CDM cosmology where $f(a) \equiv 1, g_1 = 0$.}
\label{fig:V(z)}
\end{figure}
\begin{figure}[htb]
\includegraphics[width=70mm]{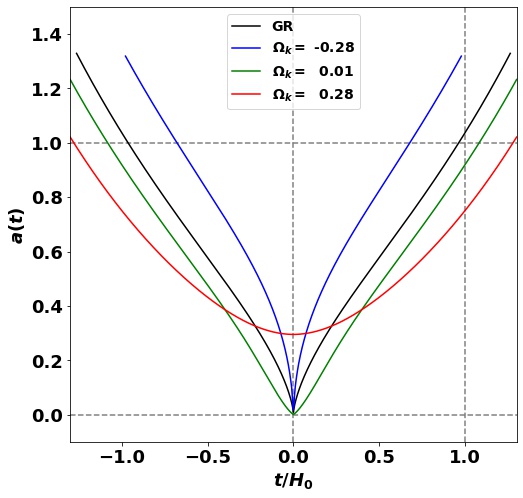}
\caption{\footnotesize The time evolution of the scale parameter from the origin to today ($a = 1$) and beyond. Notice that the Friedman equation is time- reversal invariant, such that $a(- \tau)$ denotes a deflating trajectory of the scale size. Hence the deflation to a finite bounce (Region I) or to a singularity (bang scenarios II and III) is displayed for negative conformal times.}
\label{fig:a(t)}
\end{figure}
\begin{figure}[htb]
\includegraphics[width=70mm]{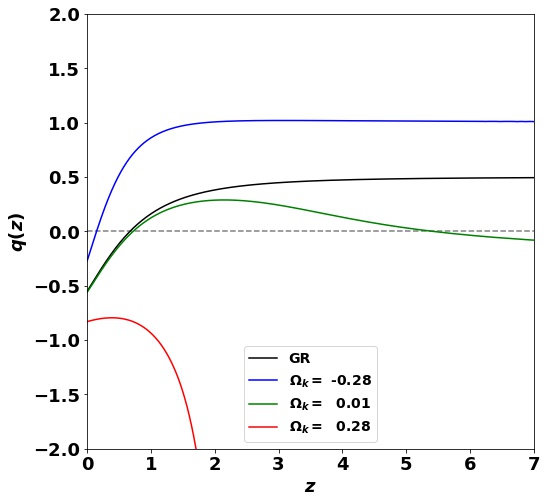}
\caption{\footnotesize The deceleration parameter $q(z)$ in the three Regions I (red), II (green) and III (blue). While in Region I a monotonically accelerating expansion is observed, accelerating and decelerating phases occur in Region II. In Region III, similarly to GR (black), an initially decelerating expansion transfers into acceleration in the dark energy era.}
\label{fig:q(z)}
\end{figure}
 %

\section{The Bounce and Bang scenarios} \label{sec:scenarioanal}

In order to get a first impression on the viability of this cosmological model we align with $\Lambda$CDM as far as possible by using the corresponding parameter set (here the ``Default'' values from Table~1) and assuming that Eq.~(\ref{eq:k0}) holds. The solution variety is seen to split up into three fundamentally different scenarios per parameter region. In \fref{fig:a(t)} the expansion trajectories for Regions I - III are plotted for typical values of the free parameter, $\Omega_K =0.28, 0.01, -0.28$ which correspond to the values of the deformation parameter $g_1 = 4.27 \times 10^{120}, -3.87 \times 10^{118}, 4.64 \times 10^{119}$, giving a considerable contribution of quadratic gravity:
\begin{itemize}
 \item \pmb{Region I} is a Bounce scenario: A deflating open universe will rapidly decelerate to a stillstand ($\dot{a}=0$) and bounce off\footnote{The CCGG bounce scenario has also been studied in~\citep{benisty18b}.} at a finite scale $a_{min}$ to proceed in a steady expansion into the dark energy era, see \fref{fig:a(t)}. The singularity is avoided due to the turning point of the corresponding potential, Eq.~(\ref{def:V(a)}), in \fref{fig:V(z)} where $V(z)/H_0^2 \equiv V(a(z))/H_0^2= \Omega_K$ is displayed using for convenience the redshift parameter $z =: 1/a-1$. The age of the universe depends on the parameter $\Omega_K>0.75\Omega_m+\Omega_r$. For the value chosen here the universe of Region I is around $30\%$ older than the popular value $H_0^{-1}$. The deceleration parameter, \fref{fig:q(z)}, is always negative and for large $z$ even $q(z) < -1$. This indicates a violation of the energy conditions and possibly an unphysical regime.
\item \pmb{Region II} wields a potential without a turning point that deviates from the potential of GR by a rather flat wide maximum, see \fref{fig:V(z)}. The evolution starts with a (Big) Bang and the scale is monotonously increasing, but in alternating acceleration and deceleration phases (\fref{fig:q(z)}). The universe can, depending on $0<\Omega_K<0.75\Omega_m+\Omega_r$ (open universe), be again significantly older than $1/H_0$ (\fref{fig:a(t)}).
\item \pmb{Region III} is comparably less spectacular. The dynamics corresponds to a slightly amended $\Lambda$CDM / Big Bang evolution. The universe is closed and consistently younger than $1/H_0$. The expansion is decelerating initially and accelerating in the late era, similarly to the GR dynamics.
\end{itemize}
The common feature of all scenarios is the graceful exit into the late dark energy era.

\section{Constraints from low-$z$ observations} \label{sec:SNeIa}

As a first test we compare the CCGG cosmology model and the standard GR $\Lambda$CDM model with the SNeIa Hubble diagram~\citep{riess04} via the formula for the extinction-corrected distance modulus, $\mu=m-M=5\log{\frac{d_{L}}{Mpc}}+25$. Thereby is
\begin{equation}\label{eq:LD1}
 d_{L}=
 \left(1+z \right) \int_0^z \frac{dz'}{H\left(z' \right)}
\end{equation}
the luminosity distance, $m$ the flux (apparent magnitude) and $M$ the luminosity (absolute magnitude) of the observed supernovae.
\begin{figure}[htb]
\includegraphics[width=70mm]{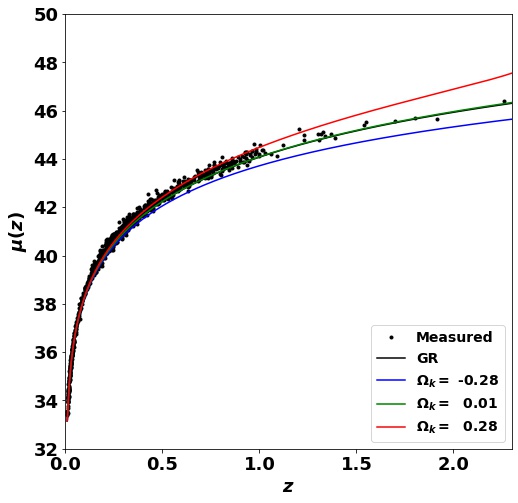}
\end{figure}
\begin{figure}[htb]
\includegraphics[width=70mm]{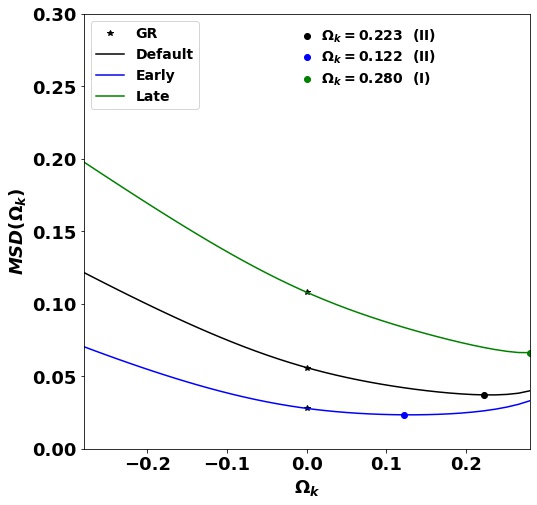}
\caption{\footnotesize The SNeIa Hubble diagrams are compared with the model prediction for the Regions I, II and III (left panel). The mean-square deviation for the two $\Lambda$~CDM parameter sets from Table 1 (right panel) display lower minima for non-zero values of the curvature parameter $\Omega_K$ than those found for standard Einstein cosmology. The minimum is found for the ``Early'' (Planck) parameter set, see Table~\ref{tab:Parameter2}.}
\label{fig:LD}
\end{figure}
The dependence of the predicted distance modulus $\mu$ on the redshift $z$ is plotted for the parameter Regions I, II, and III in the left panel of~\fref{fig:LD} and compared with the observational data.
In a sensitivity analysis w.r.t. variations of the curvature parameter, the mean-square deviation (MSD) is minimized for $\Omega_K = 0.122$, a value that points to Region II with an open geometry (see right panel of~\fref{fig:LD}).
This implies a dynamical scenario with a singular Big Bang and a secondary inflation phase\footnote{Remarkably, the ``Early'' parameter set \citep{planck18} with a smaller $H_0$ leads to a better result here even though we consider low-$z$ data. Moreover, the Hubble tension is slightly alleviated but the jury is still out on the high value of the R19 measurement. The claim that its origin is a huge local void~\citep{haslbauer20, kim20} might be an alternative explanation.}. Furthermore, with Eq.~(\ref{prooffne1}) we find $g_1 \sim 10^{119}$, i.e a significant admixture of quadratic gravity. Moreover, the fact that the relative minimum is found with the Planck parameter set indicates a potential for alleviating the so called Hubble tension.


\section{Summary and outlook}

 The key findings of the preliminary analysis presented here are:
 \begin{itemize}
  \item Torsion is identified to promote the cosmological constant to a time dependent function.
  \item The quadratic gravity term gives rise to a geometrical stress-energy with the properties of dark radiation.
  \item Solutions are consistent with the $\Lambda$CDM parameter set.
  \item All solution exit gracefully into the late dark energy era.
  \item The comparison with data suggests an open geometry and a significant admixture of Riemann-Cartan quadratic gravity in Einstein's field equations.
  \item The age of the universe can be significantly greater than $1/H_0$.
  \item After commencing with a Bang the expansion dynamics undergoes alternating acceleration and deceleration phases.
 \end{itemize}
 A comprehensive analysis  of the CCGG parameter set vs. a collection of low-$z$ data is in progress with advanced MCMC tools. Furthermore, a model for the torsion tensor that is consistent with the covariant conservation of the strain-energy tensor is under development.

 \subsection*{Acknowledgments}
This work has been supported by the Walter Greiner Gesellschaft zur F\"orderung der physikalischen Grundlagenforschung e.V., and carried out in collaboration with the Goethe University Frankfurt, Ben Gurion University of the Negev Beer Sheva, and with GSI Darmstadt. DV and JK especially thank the Fueck-Stiftung for support. We thank the following students for their contributions to the CCGG development: David Benisty, Adrian K\"onigstein, Johannes M\"unch, Dirk Kehm, Patrick Liebrich, Julia Lienert, Vladimir Denk and Leon Geiger.
We are also indebted to colleagues from the local and international community for many valuable discussions:
Prof. Horst St\"ocker (Frankfurt), Prof. Friedrich Hehl (K\"oln), Prof. Gerard ‘t Hofft (Utrecht), Prof. Eduardo Gundelmann (Beer Sheva), Prof. Peter Hess (Mexico City), Dr. Andreas Redelbach (Frankfurt), Dr. Leonid Satarov (Frankfurt), Dr. Frank Antonsen (Copenhagen),  Prof. Pavel Kroupa (Bonn), Prof. Luciano Rezolla (Frankfurt), Prof. Stefan Hofmann (M\"unchen), and Dr. Matthias Hanauske (Frankfurt).

\bibliography{IWARA_CCGG.bib}

\end{document}